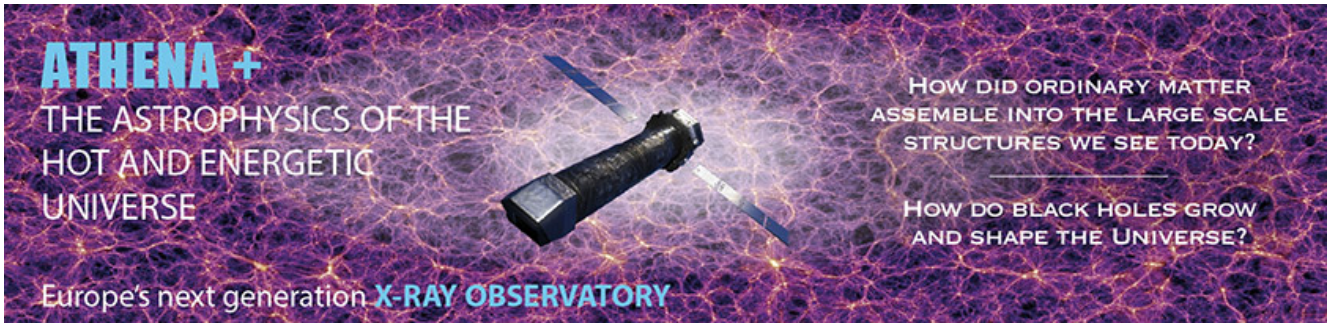

# The Hot and Energetic Universe

An *Athena+* supporting paper

# Luminous extragalactic transients

## Authors and contributors

**Peter Jonker, Paul O'Brien**, Lorenzo Amati, Jean-Luc Atteia, Sergio Campana, Phil Evans, Rob Fender, Chryssa Kouveliotou, Giuseppe Lodato, Julian Osborne, Luigi Piro, Arne Rau, Nial Tanvir, Richard Willingale

## 1. EXECUTIVE SUMMARY

The universe is dynamic and ever-changing. Even on human timescales variations can be dramatic. By studying short-timescale variations occurring in large, star-sized entities, we are probing the extremes of the parameter space of physics: extreme magnetic and/or gravitational fields accelerate particles to velocities approaching the speed of light and the strongest fields can bring matter to extreme densities and temperatures. These circumstances are naturally found around black holes of various masses – from the super-massive to stellar-mass – but can also be related to phenomena involving neutron stars, white dwarfs and collapsing massive stars. The conditions encountered in the Universe during these transient, often explosive or catastrophic events cannot be created in man-made laboratories. This realization draws the attention of physicists to the space-laboratories that Nature provides. Hence transients are prime targets for electromagnetic, gravitational wave and neutrino detectors.

X-ray observations play a crucial role in understanding and utilizing transient phenomena, which are often prodigious sources of high-energy photons. X-ray data are crucial for estimating the energy involved in the transients and for providing the diagnostics to study the physical conditions of the events and their environments. *Athena+* will provide the necessary sensitivity and spectral resolution for the study of the dynamic high-energy sky. Its peak throughput near 1 keV makes it especially well suited for sources of considerable redshift. Indeed, as the highest luminosity transients are the brightest high-redshift sources of light, the fast response capability of *Athena+* coupled with its light gathering power make it a unique facility for the study of the dynamic, evolving Universe. Gamma-ray bursts probe into the era of formation of the first stars and black holes in the Universe and *Athena+* is uniquely equipped to allow their investigation. Radio telescopes, such as the Square Kilometer Array (SKA), scanning optical telescopes, such as the Large Synoptic Survey Telescope (LSST) and the Zwicky Transient Facility and finally, multi-messenger facilities like the Cherenkov Telescope Array will provide triggers for *Athena+* including a likely classification of the transient on the basis of their observed properties. As much of the phase space of the transient sky has not yet been opened, serendipitous discoveries are virtually guaranteed. Some of the scientific goals of time-domain astronomy using *Athena+* that we can foresee now include:

- *Athena+* can provide high-quality, high-resolution X-ray spectra of bright Gamma-Ray Bursts (GRBs) at all redshifts to probe the environment of the host galaxy and track metal enrichment. The chemical fingerprint of such Pop III star explosions is distinct from that of later generations, opening the possibility to probe the IMF of the Universe. These data will likewise be searched for emission line features due to the GRB progenitor and its close environment.
- Understanding type II supernova explosions and determining which stellar systems are the progenitors of type Ia supernovae by deriving the supernova progenitor properties using *Athena+* observations of supernova shock break out and off-axis GRBs.
- Study, via tidal disruption events (TDEs), the population of dormant super-massive and potentially intermediate-mass black holes. Intermediate-mass black holes constitute a group of black holes that is invoked to explain the presence of super-massive black holes early in the Universe, but these intermediate-mass black holes have remained elusive so far.

## 2. INTRODUCTION

The universe is dynamic and ever-changing. Even on human timescales variations can be dramatic. By studying short-timescale variations occurring in large, star-sized entities, we are probing the extremes of the parameter space of physics: extreme magnetic and/or gravitational fields accelerate particles to velocities approaching the speed of light and the strongest fields can bring matter to extreme densities and temperatures. These circumstances are naturally found around black holes of various masses – from the super-massive to stellar-mass – but can also be related to phenomena involving neutron stars, white dwarfs and collapsing massive stars. The conditions encountered in the Universe during these transient, often explosive or catastrophic events cannot be created in man-made laboratories. This realization draws the attention of physicists to the space-laboratories that Nature provides. Hence transients are prime targets for electromagnetic, gravitational wave and neutrino detectors.

X-ray observations play a crucial role in understanding and utilizing transient phenomena, which are often prodigious sources of high-energy photons. X-ray data are crucial for estimating the energy involved in the transients and for providing the diagnostics to study the physical conditions of the events and their environments. *Athena+* will provide the necessary sensitivity and spectral resolution for the study of the dynamic high-energy sky. Its peak throughput near 1 keV makes it especially well-suited for sources of considerable redshift. Indeed, as the highest luminosity transients are the brightest high-redshift sources of light, the fast response capability of *Athena+* coupled with its light gathering power make it a unique facility for the study of the dynamic, evolving Universe (see Figure 1).

Time-domain astronomy will be one of the main fields in astronomy over the coming decades. The main reason why it is now finally possible to address the important physics questions by studying the extreme time-variable events in the Universe is that computational power has become sufficient to sift through huge volumes of data in real time in order to find interesting, sometimes brief, and often one-off explosive events. Examples of this capability include radio all-sky/hemisphere telescopes, such as the Square Kilometer Array (SKA), and scanning optical telescopes, such as the Large Synoptic Survey Telescope (LSST). Within a decade we expect to see the first detections of gravitational-waves and neutrinos from explosive sources. We require a major facility like *Athena+* to truly take advantage of the time-domain, multi-messenger era.

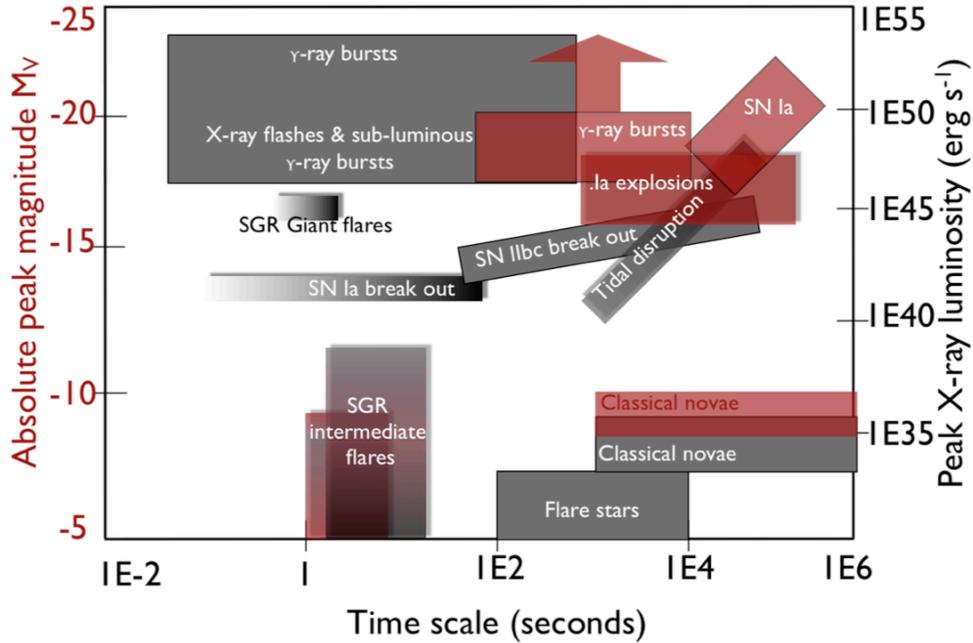

Figure 1: The absolute optical magnitude (label on the left and colours in red) or X-ray luminosity (label on the right and in black/grey) is plotted as a function of the characteristic time. Sources/phenomena that have been barely studied are plotted without border and with fading shading. So far, the lower luminosity range in this plot is not well studied, especially those sources with characteristic time scales of 1 day or less. The Athena+ effective area is such that one can search for flux variability at a signal-to-noise ratio >10 for all source classes and their fluxes observed so far in 10 seconds. Obviously, the volume we will probe with Athena+ is much larger implying that also lower flux levels can and will be observed. This capability will push the frontier for all classes of sources in the upper part of the diagram (GRBs, SNe, tidal disruption events) into the high redshift Universe.

In this supporting paper we outline some of the scientific goals of time-domain astronomy using *Athena+*, including:

- Using high quality, high-resolution X-ray spectra of bright Gamma-Ray Burst (GRB) afterglows, *Athena+* will probe the environment of GRB progenitors and their host galaxies at all redshifts. This will trace the first generation of massive stars in the Universe.
- Understanding type II supernova explosions and determining which stellar systems are the progenitors of type Ia supernovae by deriving the supernova progenitor properties using observations of supernova shock break out and off-axis GRBs. White dwarf detonation is the leading model to explain type Ia supernovae. Establishing the size of the companion star in a sample of type Ia supernovae could determine the relative fractions of progenitors among the contending models.
- Study, via bright tidal disruption events (TDEs), the population of dormant super-massive and potentially intermediate-mass black holes. Intermediate-mass black holes constitute a group of black holes that is invoked to explain the presence of super-massive black holes early in the Universe, but that have remained elusive so far.

Triggers for GRBs, Tidal disruption events as well as supernova shock break-out will come from sky monitors, including SKA, LSST, the Cherenkov Telescope Array, the Zwicky Transient Facility and any space-based electromagnetic monitors. Thus even if no X-ray or gamma-ray all-sky monitor is operational when *Athena+* flies we are confident that the events discussed here will be detected and can be follow-up timely. By 2028 multi-messenger monitoring will be routine and a large area X-ray telescope with high spectral resolution and fast slew time capability like *Athena+* will form an essential element in the follow-up observations of transient triggers. *Athena+* has a baseline design to slew within 4 hours to accepted Target of Opportunity triggers with a goal of a slew within 2 hours.

## 3. GAMMA-RAY BURSTS

### 3.1. Tracing metal enrichment and primordial stellar populations in the Universe.

GRBs can play a unique role in the study of metal enrichment through X-ray spectroscopy of the immediate environment around a burst as they are the brightest light sources at all redshifts and, for long duration events (LGRBs), occur in star-forming regions. As LGRB progenitors are short-lived massive stars, they provide an ideal probe of the effect of stellar evolution on galaxy chemical enrichment across cosmic time. X-ray spectroscopy has the unique capability of simultaneously probing all the elements (C through Ni), in all their ionization stages and all binding states (atomic, molecular, and solid), and thus provides a model-independent survey of the metals. High-resolution spectroscopy with *Athena+* would extend the frontier of such studies into the high-redshift Universe, tracing the early Population III stars and their environment. The chemical fingerprint of exploding Population III stars is distinct from that of later generations. *Athena+* X-IFU observations of GRB afterglows will determine the redshift of the GRB and typical masses of early stars. Knowledge of their masses impacts the models of the earliest stellar populations.

We adopt the Swift distribution of X-ray afterglows to provide flux levels, assume 200 GRB events will be localised per year and we conservatively assume an observation start time of 12 hours post trigger (*Athena+* has a predicted response time to a ToO of 2-4 hours). There should be around 10 GRBs per year accessible to *Athena+* with a fluence (0.3-10 keV) $\geq 10^{-6}$ erg cm$^{-2}$, which will allow 2.5 eV spectroscopy with a signal-to-noise ratio such that one can detect absorption features with an EW around 0.07 eV. A less conservative observation start time of 6 hours would substantially increase the number of GRBs for which high-quality spectra can be obtained.

Beginning with metal free (Population III) stars, the cycle of metal enrichment started when their final explosive stages injected the first elements beyond Hydrogen and Helium into their pristine surroundings, quickly enriching the gas. These ejecta created the seeds for the next generation of stars (Population II). So the cycle of cosmic chemical evolution began. Finding and mapping the earliest star formation sites (Population III/II stars) is one of the top priorities for future astrophysical observatories. Tracing the first generation of stars is crucial for understanding cosmic re-ionization, the formation of the first seed black holes (BHs), and the dissemination of the first metals in the Universe. Photons from Pop III stars and radiation generated from accretion onto the first BHs initiate the Cosmic Dawn. Later on, less massive stars and quasars sustain the high degree of ionization that renders the Universe transparent. The interplay between light and matter is strongly affected by the presence of metals. The chemical fingerprint of Pop III star explosions is distinct from that of later generations, opening the possibility to probe the IMF of the Universe. Stellar evolution studies show that the nucleosynthetic yields of Pop III and Pop II SNe differ significantly. The convolution of these yields with an IMF directly translates to abundance patterns, which can differ up to an order of magnitude depending on the characteristic mass scale of the IMF, disentangling a Population III from a Population II stellar site (Heger & Woosley 2010). Measuring these patterns using GRBs, combined with *Athena+* studies of AGN sightlines, galaxies and SNe, will enable us to determine the typical masses of early stars, thereby testing whether or not the primordial IMF is indeed top heavy.

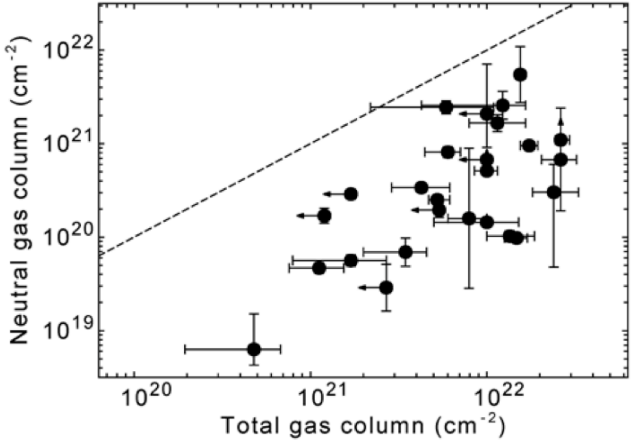
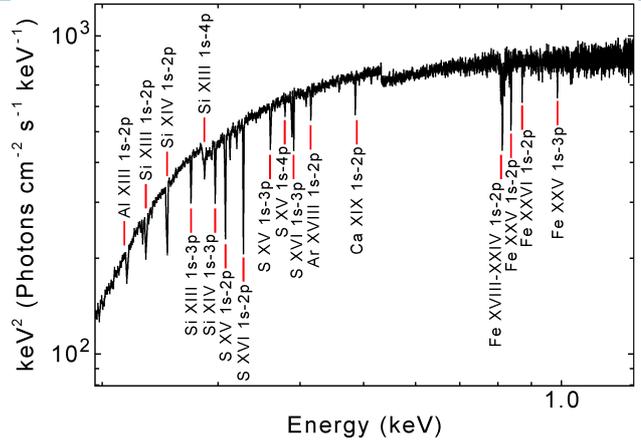

**Figure 2:** The neutral gas column density derived from optical spectroscopy compared to the total gas column density derived from X-ray observations of GRBs. The neutral column density is about 2 orders of magnitude lower, indicating that the bulk of the absorbing gas is ionized making it observable in X-rays (Schady et al. 2010).

**Figure 3:** A simulated X-IFU X-ray spectrum of a GRB afterglow at z=7, showing the capability of *Athena+* in tracing the primordial stellar populations. This medium bright afterglow (fluence=$0.4\times10^{-6}$ erg cm$^{-2}$) is characterized by deep narrow resonant lines of Fe, Si, S, Ar, Mg, from the ionized gas in the environment of the GRB. An effective column density of $2\times10^{22}$ cm$^{-2}$ has been adopted. The abundance pattern measured by *Athena+* can distinguish Population III from Population II star forming regions.

Existing data indicate that the effective column density of metals in GRB lines of sight are of the order $Z \times N_H = 10^{22}$ cm$^{-2}$, where Z is the metallicity relative to solar including the highest redshift GRBs (Starling et al., 2013). Utilizing the exquisite sensitivity and resolution of the X-IFU, *Athena+* can resolve and measure the X-ray absorption features coming from this gas, much of which is likely to be ionized and thus expected to be only accessible by observing narrow resonant X-ray absorption lines (Figure 2). The *Athena+*/X-IFU will be able to measure metal abundance patterns for a variety of ions (e.g., S, Si, Fe) for at least 10 medium-bright X-ray afterglows per year with H equivalent column densities as small as $10^{21}$ cm$^{-2}$ and gas metallicities as low as 1% of solar for the denser regions expected in early stars-forming zones; in even denser regions the accuracy will be further improved (Figure 3). This level of accuracy will discriminate between abundance patterns resulting from different nucleo-synthesis sources and the effects of a modified, non-universal IMF (Heger & Woosley 2010).

### 3.2. GRB progenitors and jet physics

Using the same spectroscopy as for the host galaxy/GRB environment studies, *Athena+* data will enable a search for emission and other spectral features due to the immediate environment around the GRB. X-ray emission line features have been suggested (e.g. Reeves et al., 2002) and could arise from the breakout of a jet cocoon or from illumination of the outer stellar envelope or pre-ejected material (Piro et al. 2000; Meszaros, 2002). These features will be temporally variable and have low equivalent width against a bright afterglow requiring confirmation using much higher throughput X-ray data than currently available.

The X-IFU data will also provide very high time resolution automatically, permitting a search for quasi-periodic oscillation (QPO) signals due to orbiting material feeding the engine, progenitor remnants (e.g. a magnetar) or for probing the jet. The accretion flow around the newly formed "central engine" is likely to be rapidly variable in structure, possibly accounting for some of the X-ray flaring behaviour commonly seen in GRBs. The formation of an accretion disk or torus could lead to a quasi-periodic signal.

The studies of GRB afterglow and progenitors can be done at all redshifts, probing a large range in GRB luminosities. The X-ray data will enable an accurate study of the relation between beamed emission (such as GRBs) and supernova explosions and it will complement data of multi-messenger observatories. Finally, GRBs are important for the study of the warm-hot intergalactic medium; this is described in detail in a separate supporting paper (Kaastra, Finoguenov et al. 2013 *Athena+* supporting paper).

The Hot and Energetic Universe: Selected transient science with Athena+## 4. TIDAL DISRUPTION EVENTS

Stellar dynamical models predict that once every $10^3$–$10^5$ year a star in a galaxy will pass within the tidal disruption radius of the central black hole and thus will be torn apart by tidal forces (Wang & Merritt 2004). The fall-back of debris onto the black hole produces a luminous electromagnetic flare that is detectable in UV and X-ray light. Indeed, a couple of UV transients coincident with the centre of a galaxy have been detected (e.g. Gezari et al., 2012). Candidates detected so far in X-rays using ROSAT, XMM-*Newton*, and *Chandra*, had black body temperatures with kT = 0.04 - 0.12 keV.

Recently, a new manifestation of tidal disruption events was reported. The new extreme source discovered by *Swift* (Swift J164449.3+573451) is probably caused by relativistic jet emission launched due to a tidal disruption event (Levan et al. 2011; Bloom et al. 2011). The source had an extreme X-ray luminosity ($10^{47}$ erg s$^{-1}$) that lasted for months. The all-sky rate for this and similar events in the volume accessible by *Swift* BAT is 2 in 7 years as a second event with similar properties was found (Swift J2058+0516; Cenko et al., 2012). The larger sensitivity for such events from LSST and radio all-sky monitors like SKA implies that *Athena+* follow-up observations will allow us to search for (redshifted) spin sensitive properties. One of the unique measurements that *Athena+* could make is when at late times X-ray absorption lines appear that could help identify the composition of the disrupted star (e.g. Strubbe & Quataert 2011) and thus determine information on its size which is important for helping constrain the black hole mass. The rate at which the SKA will find TDEs can be estimated by using the results of van Velzen et al. (2013) taking a flux limit for the SKA in survey mode of 0.1 milliJansky. From this we find that the SKA should detect more than 200 events per year. All of these are expected to be easily detectable by *Athena+* in X-rays for months if we take SwiftJ164449.3+573451 as the standard but also if we take a less extreme example such as the ROSAT detected TDEs as standard. The predictions for the LSST imply similar number of TDEs per year (more than 100 TDEs per year; Gezari et al. 2009).

Tidal disruption events have a rise time of days to weeks, the timescale for the light curve to peak depends on the black hole mass, with lower mass black holes having shorter rise times. Hence, focussing on the shorter duration events, one selects the lower-mass black hole systems. Similarly, intermediate-mass black holes will have harder X-ray spectra than more massive black holes. Intermediate-mass black holes may be remnants from the first population of massive stars, they may be formed "primordially", or they may be those black holes that grew no further from smaller seeds. Intermediate-mass black holes may comprise up to tens of per cent of the dark matter content of the Universe. A good intermediate-mass black hole candidate is HLX-1 (Farrell et al. 2009; see also the Motch, Wilms et al. 2013, *Athena+* supporting paper). Black holes with a mass less then $10^7 M_{sun}$ should produce super-Eddington flares - a regime that is currently poorly studied - at least at the beginning of the event (relevant papers describing different aspects of the super-Eddington phase are Loeb & Ulmer (1997); Strubbe & Quataert (2009), and Lodato & Rossi (2011)), while for $M_{black\ hole}$ larger than $10^7 M_{sun}$ the flare should be almost always sub-Eddington. For black hole masses larger than $10^8 M_{sun}$ we expect the star to be swallowed whole (without being disrupted first).

Dynamical studies of nearby galaxies suggest that probably all galaxies with a bulge component harbour a central supermassive black hole (SMBH). This ubiquitous presence of SMBHs is consistent with the high number density of quasars at high redshift, implying that the present SMBHs are their descendants. The bulge velocity dispersion and black hole mass appear to be tightly correlated; this is often referred to as the $M_{black\ hole}$ - $\sigma$ relation (Magorrian et al. 1998; Ferrarese & Merritt 2000). Why there is such a correlation between the black hole mass and the bulge mass is unclear as the gravitational influence sphere of the black hole is orders of magnitude too small to explain the observed correlation. Direct detection of SMBHs in the centres of quiescent galaxies is, however, limited by the need to spatially resolve the sphere of influence of the central black hole, thus limiting the searches to a small number of nearby galaxies. Events caused by the tidal disruption of stars by otherwise dormant black holes in the centres of galaxies provide a fresh look at black holes in a sample of galaxies that is independent from the current sample that is making up the $M_{black\ hole}$ - $\sigma$ relation.

The properties, such as the X-ray luminosity, its evolution with time and the spectral energy evolution of the disruption event depend on the black hole mass and spin. Establishing the spin of a larger sample of SMBHs can put the theoretical calculations about the spin alignment history of accreting SMBHs and the build-up of SMBHs via black hole mergers on a quantitative footing. When looking on short timescales one might observe dynamics (such as orbital precession of the debris) caused by spin-induced torques, thus constraining the black hole spin with a method that is not biased towards high spins. Finally, the black hole spin is an important parameter in the amplitude of the

Page 5

gravitational wave recoil kick that merging black holes are predicted to receive. If the two black holes that merge spin close to the maximal spin possible, the recoil velocity can be large enough to unbind the ensuing newly merged black hole from its parent galaxy. This process should enlarge the spread in the $M_{black\ hole}$ - $\sigma$ relation. The current sample of black holes needs to be enlarged to disentangle the importance of this effect from potential other effects such as redshift evolution of the $M_{black\ hole}$ - $\sigma$ relation.

## 5. SUPERNOVA SHOCK BREAK OUT

The best-known case of a transient serendipitously discovered in a pointed X-ray observation, is the supernova shock break out discovered by Soderberg et al. (2008). Prompt discovery of supernovae and the properties of the shock X-ray emission allow for the size of the progenitor star to be determined. This is of crucial importance as only very few supernova simulations enable the star to explode. More observational input especially during the first phase of the explosion is sorely needed to provide new constraints to the physics governing these simulations. Additional crucial input would come from the detection of the weak neutrino and gravitational wave signals that trace the explosion mechanism. However, for these signals to be detected one needs to determine the time of the explosion as accurately as possible to limit the parameter space that needs to be searched and thus limit the number of trials involved. The classical method of supernova discovery relies on their optical brightening, however, this typically only takes place several days after the explosion as the optical emission is powered mainly by the radioactive decay of the $^{56}$Ni synthesized during the supernova explosion.

The supernova shock break out marks the first escape of photons related to the explosion. The atmosphere is heated by the supernova blast wave to temperatures of $10^5$-$10^6$ K: the ensuing X-ray luminosities range from $10^{41}$ to $10^{44}$ erg s$^{-1}$. All core-collapse supernovae are expected to show a (X-ray/ultra-violet) flash upon break out. The signal lasts seconds to hours depending on the radius of the exploding star and the density of the surrounding medium (which is set by the mass-loss rate in the period leading up to the supernova explosion). So, with the ToO response time of 4 to 2 (goal) hours *Athena+* would be able to detect the longer of these events.

Following the supernova shock breakout, the early supernova light curve is powered by radiation from the expanding and cooling supernova ejecta. The luminosity and temperature evolution of the early thermal expansion phase can be used to determine the radius of the progenitor star as well as the ratio between the explosion energy and the ejecta mass. This leads to a determination of the nature of the progenitor and links the various supernova types and the progenitor stars. The detection of supernova progenitors by other means is difficult, and as such, has only been possible for a small number of nearby supernovae with pre-explosion high resolution imaging such as available from the Hubble Space Telescope.

Kasden (2010) shows that type Ia supernovae will also produce an X-ray flare similar in nature to the supernova shock breakout of core-collapse supernovae. The main idea is that the supernova ejecta interacts with the secondary, mass transferring star in the binary scenario for type Ia supernovae. The ensuing shock will heat the outer layer of the companion star such that it will emit X-rays. As the companion star rotates the heated face of the companion star will rotate in and out our line of sight producing a periodic signal. Rapid follow-up by *Athena+* of SKA radio transient alerts (SN1987A was detected within two days; Turtle et al. 1987) and/or LSST alerts of the first phase of the explosion would allow *Athena+* observations to study the crucial early phase of the SN explosion and enable us to determine the nature of the companion star in Type Ia SNe.

## 6. REFERENCES

Bloom J.S., et al., 2011, Science, 333, 203
Cenko S.B., et al., 2012, ApJ, 753, 77
Farrell S., et al., 2009, Nature, 460, 73
Ferrarese L., Merritt D., 2000, ApJ, 539, L9
Gezari S., et al., 2009, ArXiv:0903.1107
Gezari S., et al., 2012, Nature, 485, 217
Heger A., Woosley S.E., 2010, ApJ, 724, 341
Kaastra J., Finoguenov A., et al., 2013 Athena+ supporting paper, http://www.the-athena-x-ray-observatory.eu
Kasden D., 2010, ApJ, 708, 1025
Levan A.J., et al., 2011, Science, 333, 199
Lodato G., Rossi E.M., 2011, MNRAS, 410, 359
Loeb A., Ulmer, A., 1997, ApJ, 489, L573

Magorrian J., et al., 1998, AJ, 115, 2285
Meszaros P., 2002, Ann Rev., 40, 137
Motch C., Wilms J., et al., 2013, Athena+ supporting paper, http://www.the-athena-x-ray-observatory.eu
Piro L., et al., 2000, Science, 290, 955
Reeves J., et al., 2002, Nature, 416, 512
Soderberg A., et al., 2008; Nature, 453, 469
Starling R., et al., 2013, MNRAS, in press, ArXiv: 1303:0844
Strubbe L., Quataert E., 2009, MNRAS, 400, 2070
Strubbe L., Quataert E., 2011, MNRAS, 415, 168
Turtle A.J., et al., 1987, Nature, 327, 38
van Velzen S., et al., 2013, A&A, 552, 5
Wang J., Merritt D., 2004, ApJ, 600, 149